\documentclass{webofc}

\usepackage[varg]{txfonts}   
\usepackage{hyperref}
\usepackage{url}
\DeclareMathOperator{\Tr}{Tr}
\usepackage{cleveref}
\usepackage{siunitx}
\usepackage{bm}

\hypersetup{colorlinks=true,citecolor=blue,urlcolor=blue,linkcolor=blue}
\begin{document}
\title{Prospects of measuring quantum entanglement in $\tau\tau$ final states at a future $e^+e^-$ Higgs factory}
%
%

\author{\firstname{Cedric} \lastname{Breuning}\inst{1}\fnsep\thanks{\email{ccbreuning@uni-bonn.de}} \and
        \firstname{Philip} \lastname{Bechtle}\inst{1} \and
        \firstname{Klaus} \lastname{Desch}\inst{1} \and
        \firstname{Christian} \lastname{Grefe}\inst{1}
}

\institute{Physikalisches Institut, Rheinische Friedrich-Wilhelms Universit\"at Bonn, Nussallee 12, 53115 Bonn, Germany
          }

\abstract{We introduce a method to study quantum entanglement at a future $e^+e^-$ Higgs factory (here the Future Circular Collider colliding $e^+$ and $e^-$ (FCC-ee) operating at $\sqrt{s}=\SI{240}{\giga\electronvolt}$) in the $\tau\tau$ final state. This method is focused on the $\tau \to \pi \nu_\tau$ decay. We show how the introduced method works on simulated events without detector effects. When detector effects are applied, the necessary $\tau$ four-momenta can be reconstructed from kinematic constraints. 
We will discuss the advantages of $e^+e^-$ collisions over $pp$ collisions where the reconstruction of the $\tau\tau$ rest frame is more difficult. This discussion will focus on the influence of $p_T$ trigger cuts on the visible $\pi^\pm$ in the $\tau$ lepton decay.
}
\maketitle
\section{Introduction}
\label{intro}

Einstein, Podolsky and Rosen (EPR) argued in 1935 that physical reality is not completely described by quantum mechanics (QM)~\cite{epr}. This argument, also known as EPR paradox, lead to the proposals of local hidden variable theories (LHVT) with a deterministic structure instead of the statistical approach of QM~\cite{genovese, dreiner}. Using the variant of the EPR paradox introduced by Bohm and Aharonov~\cite{bohm}, Bell found a way to test LHVT against QM and entangled states~\cite{bell}. Clauser, Horne, Shimony and Holt (CHSH) generalized Bell's theorem for realizable experiments~\cite{chsh}. Using the CHSH and extended arguments~\cite{ch}, different experiments over the last decades use photons at low energies to rule out LHVT, e.g.~\cite{freedman,aspect}. Recent studies~\cite{fabbrichesi, altakach} motivate measurements at higher energies using massive particles.

We perform a CHSH test in the $H\to\tau^+\tau^-$ process. This process is an excellent probe for this test due to the scalar nature of the Higgs boson and since the $\tau$ spin is accessible through the measurement of the $\tau$ decay products. We introduce an observable that is accessible at colliders and gives an equivalent condition as the one introduced by CHSH in~\cite{chsh}. The goal of this study is to calculate the sensitivity of this measurement at the planned FCC-ee~\cite{fcc} electron-positron collider using a fast detector simulation of the International Detector for Electron-positron Accelerators (IDEA)~\cite{fcc} using the \texttt{DELPHES~3}~\cite{delphes} fast simulation package and including the relevant background processes. We will show how this measurement works without detector effects. Then, the reconstruction of the relevant variables after detector effects will be discussed. In the end we will discuss some advantages an $e^+e^-$ collider has compared to a hadron collider. 

\section{Methods}
\label{methods}

The state of the $\tau^+\tau^-$ system can be represented by the hermitian, normalized density matrix
\begin{equation}
    \label{eq:density}
    \rho = \frac{1}{4}\left[I \otimes I + \sum_i B_i^+ (\sigma_i \otimes I)+ \sum_j B_j^- (I\otimes \sigma_j ) + \sum_{ij}C_{ij}(\sigma_i\otimes\sigma_j)\right]
\end{equation}
with the two dimensional unit matrix $I$, the Pauli matrices $\sigma_i$, the polarization of the $\tau$ leptons $B^+_i, B^-_j$ and the correlation of the $\tau$ leptons $C_{ij}$. The indices $i,j$ run over the axes of the used three dimensional orthonormal coordinate system~\cite{fabbrichesi,horodecki}. The same coordinate system $\{\bm{\hat{r}},\bm{\hat{n}},\bm{\hat{k}}\}$ as in~\cite{altakach, fabbrichesi} is chosen: the first axis $\bm{\hat{k}} = \bm{\hat{p}}_{\tau^-}$ is the flight direction of the $\tau^-$ in the $\tau\tau$ rest frame. Using ${\bm{\hat{p}} = (0,0,1)^T}$, the direction of one of the $e^\pm$ beams, the other axes are constructed as ${\bm{\hat{r}} = (\bm{\hat{p}}-\bm{\hat{k}}\cos\Theta)}/\sin\Theta$ and $\bm{\hat{n}}=\bm{\hat{k}}\times\bm{\hat{r}}$ with $\cos\Theta = \bm{\hat{k}}\cdot \bm{\hat{p}}$.

For the measurement, the observable $\mathfrak{m}_{12} = m_1 + m_2$ is constructed from the symmetric matrix  $M = C^TC$ with eigenvalues ${m_1 \geq m_2 \geq m_3}$~\cite{fabbrichesi}. Using the argument in~\cite{horodecki} it is sufficient to test
\begin{equation}
    \label{m12_condition}
    \mathfrak{m}_{12} > 1
\end{equation}
to show the violation of the inequality introduced by CHSH~\cite{chsh} and rule out LHVT. Thus, it is sufficient to measure the correlation matrix $C$ to test the CHSH inequality. $C$ can be calculated as the expectation value ${C_{ij} = \Tr(\rho(\sigma_i\otimes\sigma_j)) = \langle s_i^+s_j^- \rangle}$ where $s_i^\pm$ is the operator for the spin component in direction $\hat{\bm{i}} \in \{\bm{\hat{r}},\bm{\hat{n}},\bm{\hat{k}}\}$ of the $\tau^\pm$. Note that the spin operators are scaled by $2/\hbar$ leading to eigenvalues of $\pm 1$~\cite{altakach, fabbrichesi}.

The spin direction of the $\tau$ leptons is not measured directly in the experiment. However, information on the polarization of the $\tau$ leptons is available through the decay products. In the decay $\tau \to \pi \nu_\tau$ (1 prong and 0 neutral tracks, 1p0n) the direction of the $\pi$ allows conclusions on the $\tau$ polarization.  Assuming the $\tau^-$ is polarized in direction $\bm{\hat{s}}$, the probability that the $\pi^-$ in the 1p0n decay in the $\tau^-$ rest frame is emitted in direction $\bm{\hat{p}}_{\pi^-}$ ($\vert\bm{\hat{s}}\vert = \vert\bm{\hat{p}}_{\pi^-}\vert = 1$) is given as
\begin{equation}
    \label{piemit}
    P(\bm{\hat{p}}_{\pi^-}\vert \bm{\hat{s}}) = 1 + \alpha_{\tau^-}  \bm{\hat{s}} \cdot \bm{\hat{p}}_{\pi^-}
\end{equation}
with the spin analyzing power $\alpha_{\tau^-} = -\alpha_{\tau^+}$~\cite{altakach, bullock}. Defining $\cos \theta_i^{\pi^\pm} = \bm{\hat{p}}_{\pi^\pm} \cdot \bm{\hat{i}}$ in the $\tau^\pm$ rest frame and using \cref{piemit} it is shown in~\cite{altakach} that 
\begin{equation}
\label{eq:spincos}
    \langle \cos\theta_i^{\pi^+}\cos\theta_j^{\pi^-}\rangle = -\frac{1}{9} \langle s_i^+s_j^- \rangle\,.
\end{equation}
In a collider experiment $\bm{\hat{p}_{\pi^\pm}}$ can be measured and in an $e^+e^-$ collision it is possible to reconstruct $\bm{\hat{p}}_{\tau^\pm}$ (see \cref{sec:reco}), which is needed to calculate the coordinate axes $\bm{\hat{i}}$ and for boosts in the relevant rest frames. Using \cref{eq:spincos} the correlation matrix can be calculated with
\begin{equation}
    \label{eq:corrmatrixcalc}
     C_{ij} = -9 \cdot \langle \cos\theta_i^{\pi^+}\cos\theta_j^{\pi^-}\rangle = -9 \int \mathrm{d}\cos\theta_i^{\pi^+}\mathrm{d}\cos\theta_j^{\pi^-} \frac{\mathrm{d}\sigma \cdot \sigma^{-1}}{\mathrm{d}\cos\theta_i^{\pi^+}\mathrm{d}\cos\theta_j^{\pi^-}}\cos\theta_i^{\pi^+}\cos\theta_j^{\pi^-}
\end{equation}
with the cross section $\sigma$.

Abel, Dittmar and Dreiner point out in~\cite{dreiner} that with this approach only a subclass, of unknown size, of LHVT can be tested against QM. They claim that for an observable that is constructed from variables with commuting components, e.g. $\bm{\hat{p}}_{\pi^\pm}$, a LHVT can be constructed that produces the same results as QM. The definition of $\mathfrak{m}_{12}$ uses non-commuting spin operators. However, to create a measurable observable $P(\bm{\hat{p}}_{\pi^-}\vert \bm{\hat{s}})$ in \cref{piemit} is assumed, which is a result from QM. With this assumption the sensitivity to all LHVT is lost that predict a different $P(\bm{\hat{p}}_{\pi^-}\vert \bm{\hat{s}})$.

\section{Event Generation}
\label{sev:eventgen}

Events are generated using \texttt{MadGraph5\_aMC@NLO}~(v.3.5.3)~\cite{madgraph}. Showering,  hadronization, and the $\tau$ decay, is done using \texttt{Pythia 8}~(v.8.306)~\cite{pythia}. Fast detector simulation is performed with \texttt{DELPHES 3}~(v.3.5.1pre10)~\cite{delphes} with the IDEA configuration as implemented in the used release. The simulated process at $\sqrt{s}=\SI{240}{\giga\electronvolt}$ is
\begin{equation*}
    e^+e^- \to ZH, Z\to x\overline{x}, H\to\tau^+\tau^- \quad\text{where}\quad x\in\{u,d,s,c,b,e^-,\mu^-\},\,\overline{x}\in\{\overline{u}, \overline{d}, \overline{s}, \overline{c}, \overline{b},e^+,\mu^+\}.
\end{equation*}
It contains the 1p0n decay of the $\tau$ leptons, the most sensitive decay, among the other $\tau$ decays~\cite{altakach}. The $ZH$ production cross section at $\sqrt{s}=\SI{240}{\giga\electronvolt}$ at FCC-ee amounts to approximately \SI{200}{\femto\barn}. If \SI{5}{\per\atto\barn} of data is collected there will be approximately $10^6$ $ZH$ events~\cite{fcc}. Using the branching ratios of the $Z$ and $H$ decay, \num{46000} events of the process above are expected~\cite{pdg}.
\section{Results without detector effects}
\label{sec:restruth}

\begin{figure*}
\centering
\vspace*{1cm}   
\includegraphics[width=12cm,clip]{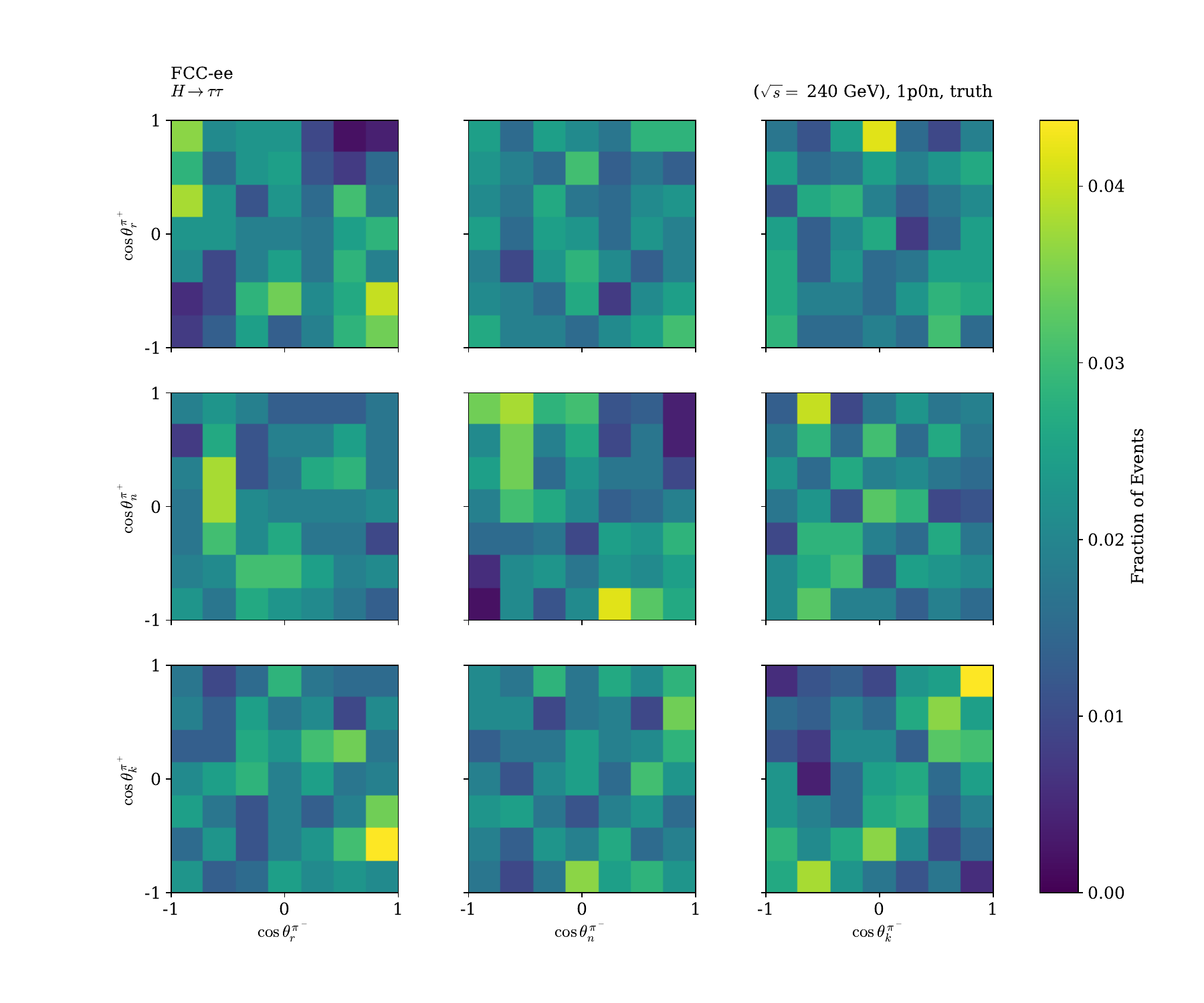}
\caption{The fraction of events differential in $\cos \theta_i^{\pi^+}$ and $\cos \theta_j^{\pi^-}$ for all axes combinations are shown. Simulated events without detector effects are used. The simulated process is described in \cref{sev:eventgen}.}
\label{fig:truthcij}       
\end{figure*}

The expected correlation matrix in the standard model is $C = \mathrm{diag}(1,1,-1)$ which leads to $\mathfrak{m}_{12} = 2$, thus violating the CHSH inequality~\cite{fabbrichesi,altakach}. The integral in \cref{eq:corrmatrixcalc} can be calculated as a sum over a two dimensional histogram of fraction of events, where every bin is multiplied by its central value. The two dimensional histograms are shown in \cref{fig:truthcij}. The diagonal elements have the expected structure, while the off diagonal elements look random. This shows in the resulting correlation matrix
\begin{equation}
    \label{resultcij}
    C = 
 \begin{pmatrix}
0.90 & 0.05 & 0.01 \\ 
-0.09 & 0.96 & -0.06 \\ 
0.18 & -0.01 & -0.94 \\ 
\end{pmatrix} 
\end{equation}
with non-zero values on the diagonal and more close to zero elements on the off-diagonal. We find $\mathfrak{m}_{12} = 1.96$ for the CHSH test. The simulated results without detector effects fit well to expectation, showing that the QM physics is appropriately implemented. However, uncertainties still have to be considered. 

\section{Reconstruction of $p_{\tau^\pm}$}
\label{sec:reco}
\Cref{sec:restruth} and also~\cite{altakach} show that this measurement works in principle. The next step is to estimate the sensitivity of this measurement in a future experiment. In this case the detector is simulated with \texttt{DELPHES 3}~\cite{delphes}. The study of data including detector effects from the fast detector simulation is still ongoing and results are not available yet. Thus, only the required steps to reconstruct all needed variables will be described here.

For the measurement the $\tau^\pm$ momenta are essential, since the $\tau\tau$ rest frame is needed. The four-momenta $p_{\tau^\pm}$ are not measured directly in the detector and have to be reconstructed. In the experiment the four-momenta of the $\pi^\pm$ of the 1p0n decay can be measured. Also the four-momentum of the $Z$ can be measured from its decay products. The four-momentum of the colliding $e^+e^-$ pair $p_\mathrm{in}$ is known (assuming initial state radiation effects can be corrected for) so the $H$ four-momentum can be calculated as $p_H = p_\mathrm{in} - p_Z$. Using the eight constraints ${p_{\tau^+} + p_{\tau^-} = p_H}$,  ${p_{\tau^\pm}^2 = m_\tau^2}$ and ${(p_{\tau^\pm} - p_{\pi^\pm})^2 = m_\nu^2 = 0}$ in the $H$ rest frame a system of nonlinear equations can be constructed which can be solved for the four-momentum components of both $\tau$-leptons. This calculation is shown in~\cite{altakach}. Since this system is nonlinear, there are two solutions to this problem. A geometrical and a lifetime argument are used to select the correct solution, detailed in the following. The $\tau^\pm$ track and the corresponding $\pi^\pm$ track are approximated as lines going in the momentum direction of the particles. The $\tau^\pm$ track originates from $(0,0,0)^T$ and the $\pi^\pm$ track from some position $\bm{x}_{\pi^\pm}$ which is extracted from the data. By solving
\begin{equation}
    \label{eq:select}
    \bm{x}_{\pi^\pm} \cdot t_{\pi^\pm} + (\bm{p}_{\pi^\pm} \times \bm{p}_{\tau^\pm}) \cdot t_{d^\pm} - \bm{p}_{\tau^\pm} \cdot t_{\tau^\pm} = \begin{pmatrix}
        0 \\ 0 \\ 0
    \end{pmatrix}
\end{equation}
for both $\tau$-leptons the closest distance between the tracks, $d = \vert(\bm{p}_{\pi^\pm} \times \bm{p}_{\tau^\pm}) \cdot t_{d^\pm}\vert$, and the length of the $\tau$ track, $l = \vert\bm{p}_{\tau^\pm} \cdot t_{\tau^\pm}\vert$ can be estimated. These two variables are calculated for both solutions $\alpha \in \{1,2\}$ for both $\tau$-leptons and for each $\tau$ the solution with smaller
\begin{equation}
    \label{eq:logl}
    -\log\mathcal{L}_\alpha = \log l_{\tau,\alpha} + \frac{l_\alpha}{l_{\tau,\alpha}} + \frac{d_\alpha^2}{\sigma_d^2}
\end{equation}
is chosen. The first two terms are the probability that the $\tau^\pm$ did not decay yet with $l_\tau = c\tau_\tau\beta\gamma$ with $c\tau_\tau = \SI{87.03}{\micro\metre}$~\cite{pdg}. Additionally, solutions with negative $l_\alpha$ are discarded because the $\tau^\pm$ would have traveled in the wrong direction. The second term, with the resolution $\sigma_d$, implements the geometrical argument that the $\tau^\pm$ track should cross the $\pi^\pm$ track, so the solution with a smaller distance between the tracks is favored. Instead of $\bm{x}_{\pi^\pm}$  the impact parameter could potentially be used for this selection.

\section{Comparison of $e^+e^-$ and $pp$ collisions}
\label{sec:comp}
The method, introduced in \cref{methods} also works for other collisions, like $pp$ collisions at the Large Hadron Collider (LHC). However, since the initial state of the process is not as well known as in an $e^+e^-$ collision, the reconstruction introduced in \cref{sec:reco} does not work. This makes it much harder to reconstruct the necessary $\tau\tau$ rest frame in a $pp$ collision. Another problem at $pp$ collisions are trigger acceptance cuts, especially on the visible transverse momentum $p_{T,\mathrm{vis}}$ which require, for example for the ATLAS trigger, for the relevant process at least \SI{40}{\giga\electronvolt} for the leading and \SI{30}{\giga\electronvolt} for the subleading $\tau$~\cite{atlastrigger}. Calculating $C_{ij}$ with \cref{eq:corrmatrixcalc} assumes no acceptance cuts~\cite{fabbrichesi}. To show the effect of the $p_{T,\mathrm{vis}}$ cuts a vector boson fusion $H\to\tau\tau$ sample at $\sqrt{s}=\SI{13}{\tera\electronvolt}$ has been generated using \texttt{MadGraph5\_aMC@NLO}~\cite{madgraph}. The $\tau$ lepton decays are handled with the \texttt{TauDecay} package~\cite{taudecay}. 

\begin{figure*}
\centering
\vspace*{1cm}       
\includegraphics[width=6cm,clip]{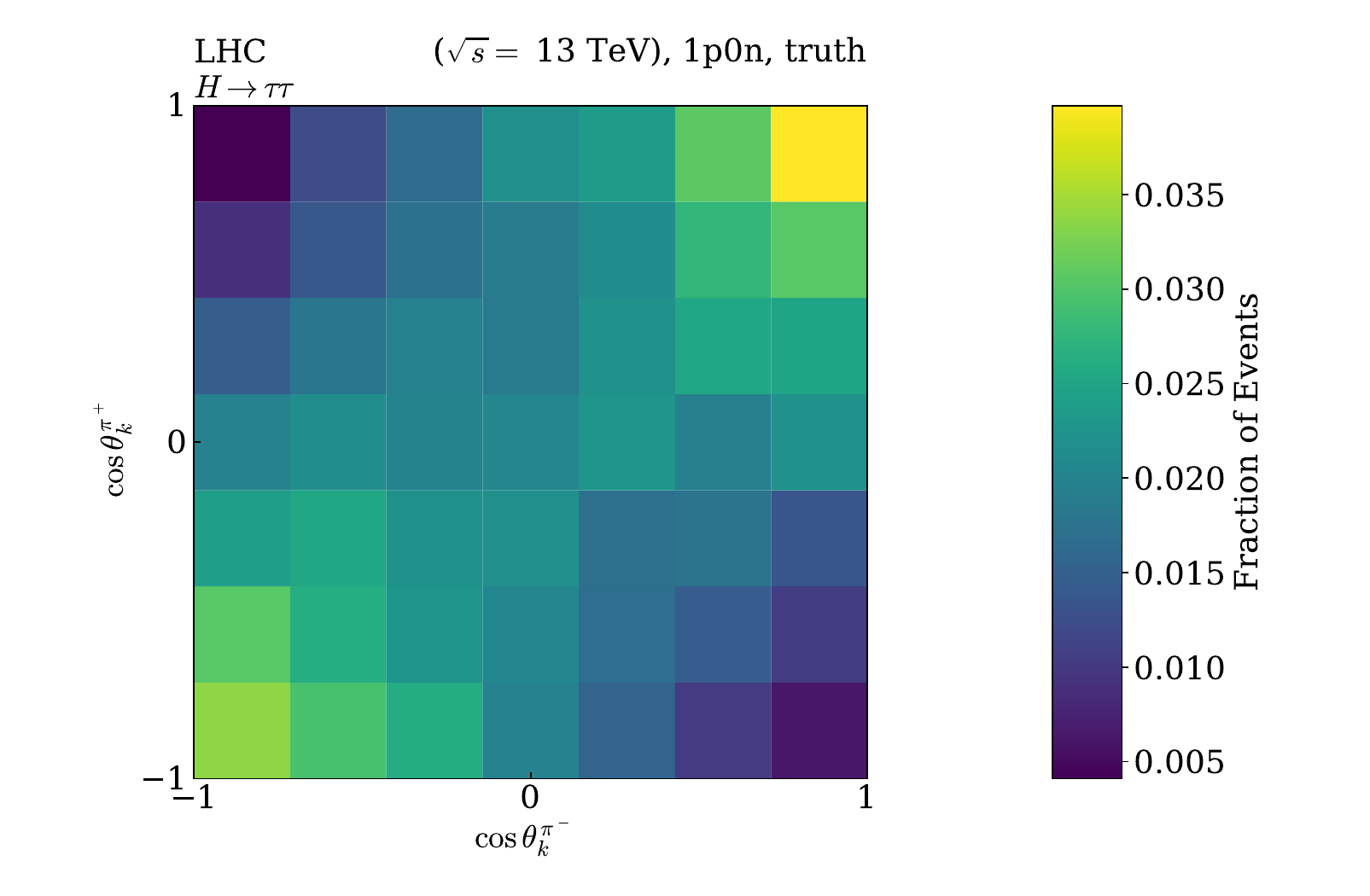}
\includegraphics[width=6cm,clip]{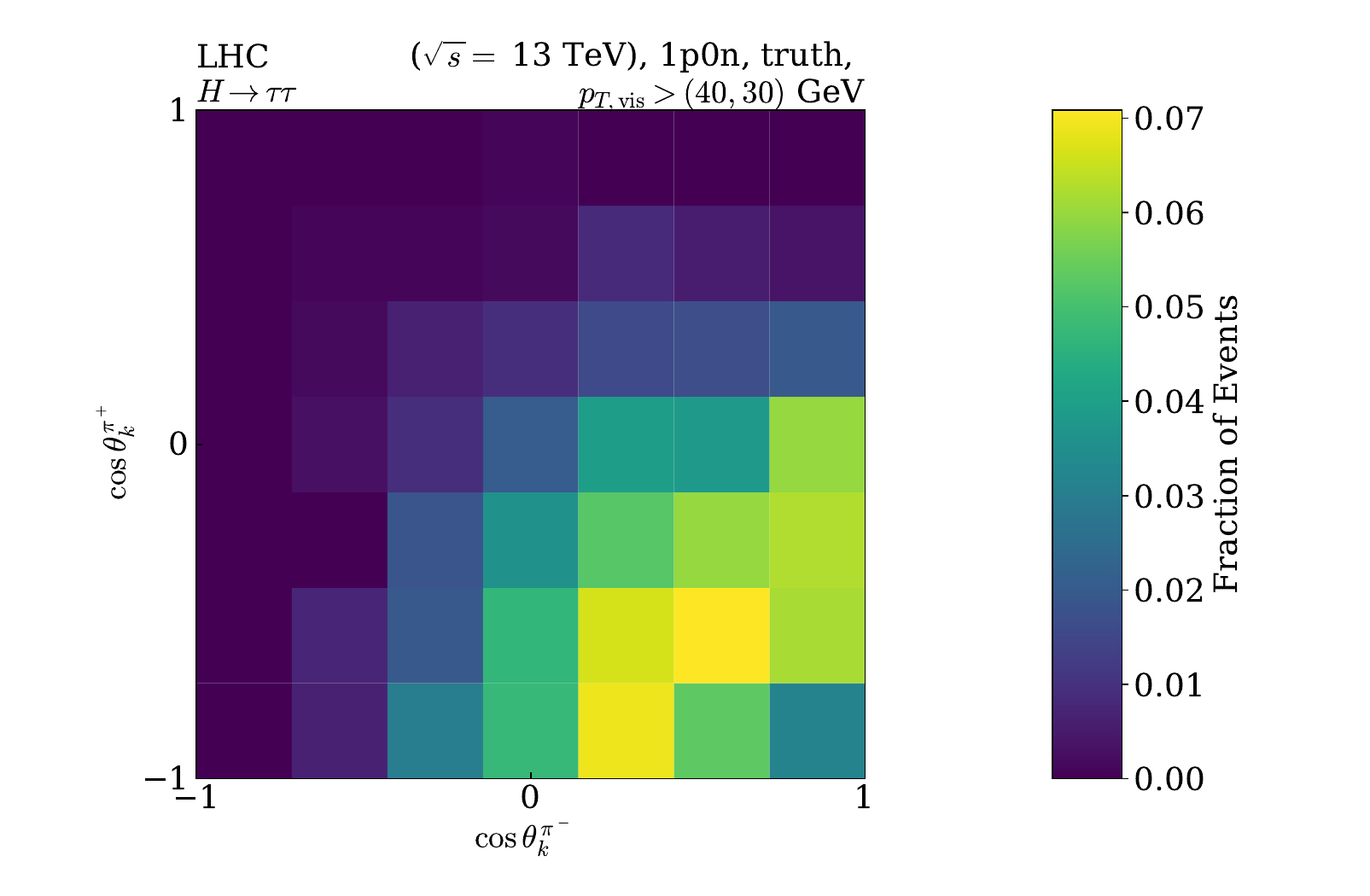}
\caption{The fraction of events differential in $\cos \theta_k^{\pi^+}$ and $\cos \theta_k^{\pi^-}$. Simulated events without detector effects are used. The simulated process is vector boson fusion $H\to\tau\tau$ in $pp$ collisions at $\sqrt{s} = \SI{13}{\tera\electronvolt}$. On the left no acceptance cuts are applied, on the right a $p_{T,\mathrm{vis}} > \SI{40}{\giga\electronvolt}$ cut on the leading $\tau$ and a $p_{T,\mathrm{vis}} > \SI{30}{\giga\electronvolt}$ cut on the subleading $\tau$ are applied.}
\label{fig-2}       
\end{figure*}

The $p_{T,\mathrm{vis}}$ cuts affect the shape of the two dimensional histograms, that are used to calculate $C_{ij}$ with \cref{eq:corrmatrixcalc}, for the $k$ components of the matrix. This can be seen in \cref{fig-2} for $C_{kk}$. This leads to positive, close to $1$, values for $C_{kk}$, which is expected to be $-1$. It may not be impossible to overcome this problem. However, the efficiency (fraction of events that survive the cut) goes to zero in the relevant regions ($\cos\theta_k^{\pi^+}\to1$, $\cos\theta_k^{\pi^-}\to1$ and $\cos\theta_k^{\pi^+}\to-1$, $\cos\theta_k^{\pi^-}\to-1$) which makes this problem even harder so solve.

Since the kinematics are much easier in $e^+e^-$ collisions, the measurement of quantum entanglement should be easier to implement compared to $pp$ collisions. The sensitivity that can be achieved at an $e^+e^-$ collider has yet to be determined.

\section{Conclusion}
\label{sec:concl}

We showed a method that would allow the measurement of quantum entanglement at a collider. The results without detector effects match the expectation, but the uncertainties have not yet been determined. We showed how the relevant variables could be reconstructed in an $e^+e^-$ collision. We also discussed the advantages $e^+e^-$ collisions would have compared to $pp$ collisions. In the next steps detector effects and background processes should be included to calculate the sensitivity this measurement could reach at an $e^+e^-$ Higgs factory. Similarly, this measurement can be performed for $Z\to\tau\tau$ at $\sqrt{s}=m_Z$~\cite{fabbrichesi}. Relevant background processes and detector effects also have to be included in the sensitivity calculation of this measurement.

%

\begin{thebibliography}{}
%
%
\bibitem{epr}
A. Einstein, B. Podolsky and N. Rosen, Can Quantum-Mechanical Description of Physical Reality Be Considered Complete?. Phys. Rev. \textbf{47}, 777-780 (1935). \url{https://doi.org/10.1103/PhysRev.47.777}
\bibitem{dreiner}
S. A. Abel, M. Dittmar and H. Dreiner, Testing locality at colliders via Bell's inequality?. Phys. Lett. B \textbf{280}, 304-312 (1992). \url{https://doi.org/10.1016/0370-2693(92)90071-B}
\bibitem{genovese}
M. Genovese, Research on hidden variable theories: A review of recent progresses. Phys. Rep. \textbf{413} (6), 319-396 (2005). \url{https://doi.org/10.1016/j.physrep.2005.03.003}
\bibitem{bohm}
D. Bohm and Y. Aharonov, Discussion of experimental proof for the paradox of Einstein, Rosen, and Podolsky. Phys. Rev. \textbf{108} (4), 1070-1076 (1957). \url{https://doi.org/10.1103/PhysRev.108.1070}
\bibitem{bell}
J. S. Bell, On the Einstein Podolsky Rosen paradox. Physics \textbf{1} (3), 195-200 (1964).  \url{https://doi.org/10.1103/PhysicsPhysiqueFizika.1.195}
\bibitem{chsh}
J. F. Clauser, M. A. Horne, A. Shimony and R. A. Holt, Proposed experiment to test local hidden-variable theories. Phys. Rev. Lett. \textbf{23} (15), 880-884 (1969). \url{https://doi.org/10.1103/PhysRevLett.23.880}
\bibitem{ch}
J. F. Clauser and M. A. Horne, Experimental consequences of objective local theories. Phys. Rev. D \textbf{10} (2), 526-535 (1974). \url{https://doi.org/10.1103/PhysRevD.10.526}
\bibitem{freedman}
S. J. Freedman and J. F. Clauser, Experimental test of local hidden-variable theories. Phys. Rev. Lett. \textbf{28} (14), 938-941 (1972). \url{https://doi.org/10.1103/PhysRevLett.28.938} 
\bibitem{aspect}
A. Aspect, J. Dalibard and G. Roger, Experimental test of Bell's inequalities using time-varying analyzers. Phys. Rev. Lett. \textbf{49} (25), 1804-1807 (1982). \url{htx`tps://doi.org/10.1103/PhysRevLett.49.1804}
\bibitem{fabbrichesi}
M. Fabbrichesi, R. Floreanini and E. Gabrielli, Constraining new physics in entangled two-qubit systems: top-quark, tau-lepton and photon pairs. Eur. Phys. J.C. \textbf{83}, 162 (2023). \url{https://doi.org/10.1140/epjc/s10052-023-11307-2}
\bibitem{altakach}
M. M. Altakach, P. Lamba, F. Maltoni, K. Mawatari and K. Sakurai, Quantum information and $CP$ measurement in $H\to\tau^-\tau^+$ at future lepton colliders. Phys. Rev. D \textbf{107} (9), 093002 (2023). \url{https://doi.org/10.1103/PhysRevD.107.093002}
\bibitem{fcc}
FCC collaboration, FCC-ee: The Lepton Collider: Future Circular Collider Conceptual Design Report Volume 2. Eur. Phys. J. Special Topics \textbf{228} (2), 261-623 (2019). \url{https://doi.org/10.1140/epjst/e2019-900045-4}
\bibitem{delphes}
The DELPHES 3 collaboration, J. de Favereau, C. Delaere, P. Demin, A. Giammanco, V. Lema\^itre, A. Mertens and M. Selvaggi, DELPHES 3: a modular framework for fast simulation of a generic collider experiment.  J. High Energ. Phys. \textbf{2014}, 57 (2014). \url{https://doi.org/10.1007/JHEP02(2014)057}. \href{https://arxiv.org/abs/1307.6346}{arXiv:1307.6346} \textbf{[hep-ex]}
\bibitem{horodecki}
R. Horodecki, P. Horodecki and M. Horodecki, Violating Bell inequality by mixed spin-$\frac{1}{2}$ states: necessary and sufficient condition. Phys. Lett. A \textbf{200} (5), 340-344 (1995). \url{https://doi.org/10.1016/0375-9601(95)00214-N}
\bibitem{bullock}
B. K. Bullock, K. Hagiwara and A. D. Martin, Tau polarization and its correaltions as a probe of new physics. Nuc. Phys. B \textbf{395} (3), 499-533 . \url{https://doi.org/10.1016/0550-3213(93)90045-Q}
\bibitem{madgraph}
J. Alwall, R. Frederix, S. Frixione, V. Hirschi, F. Maltoni, O. Mattelaer, H.-S. Shao, T. Stelzer, P. Torrielli and M. Zaro, The automated computation of tree-level and next-to-leading order differential cross sections, and their matching to parton shower simulations.  J. High Energ. Phys. \textbf{2014} (7), 79 (2014). \url{https://doi.org/10.1007/JHEP07(2014)079}. arXiv: \href{https://arxiv.org/abs/1405.0301}{arXiv:1405.0301} \textbf{[hep-ph]}
\bibitem{pythia}
C. Bierlich, S. Chakraborty, N. Desai, L. Gellersen, I. Helenius, P. Ilten, L. L\"onnblad, S. Mrenna, S. Prestel, C. T. Preuss, T. Sj\"ostrand, P. Skands, M. Utheim and R. Verheyen, A comprehensive guide to the physics and usage of PYTHIA 8.3. SciPost Phys Codebases \textbf{8} (2022). \url{https://doi.org/10.21468/SciPostPhysCodeb.8}. \href{https://arxiv.org/abs/2203.11601}{arXiv:2203.11601} \textbf{[hep-ph]}
\bibitem{pdg}
S. Navas et al. (Particle Data Group), Phys. Rev. D \textbf{110}, 030001 (2024). \url{https://doi.org/10.1103/PhysRevD.110.030001}
\bibitem{atlastrigger}
ATLAS Collaboration, The ATLAS Tau Trigger in Run 2. ATLAS-CONF-2017, \textit{ATLAS-CONF-2017-061} (2017)
\bibitem{taudecay}
K. Hagiwara, T. Li, K. Mawatari and J. Nakamura, TauDecay: a library to simulate polarized tau decays via FeynRules and MadGraph5. Eur. Phys. J. C \textbf{73}, 2489 (2013). \url{https://doi.org/10.1140/epjc/s10052-013-2489-4}
\end{thebibliography}
%
%

\end{document}